\documentclass[a4paper]{jpconf}

\usepackage{hyperref}
\usepackage{amsfonts,amssymb,amsmath,mathtools} 
\usepackage{graphicx}
\usepackage[symbol*]{footmisc}
\usepackage{xcolor}

\usepackage{caption}
\usepackage{subcaption}

\newtheorem{proposition}{Proposition}[section]
\newtheorem{theorem}[proposition]{Theorem}

\newtheorem{definition}[proposition]{Definition}
\newcommand {\proof} {\par\textit{Proof}. \ignorespaces}
\newcommand {\eproof}
      {\null\hfill{\large$\Box$}}

\newcommand{\vektor}[1]{{\boldsymbol{#1}}}

\newcommand{\vx}{\mathbf{x}}
\newcommand{\vy}{\vektor{y}}
\newcommand{\vz}{\vektor{z}}
\newcommand{\vPhi}{\vektor{\Phi}}

\DefineFNsymbols*{lamport}{*{\textsuperscript{\textdagger}}\ddagger\S\P\|%
{**}{\dagger\dagger}{\ddagger\ddagger}}

\setfnsymbol{lamport}

\begin{document}


{\small \hspace{-5mm}HU-EP-12/46\\SFB/CPP-12-88\\DESY 12-211}

\title{A first look at quasi-Monte Carlo\\for lattice field theory problems}

\renewcommand{\thefootnote}{\fnsymbol{footnote}}

\author{K.~Jansen\textsuperscript{1}, H.~Leovey\textsuperscript{2}, A.~Nube\textsuperscript{1,3}, 
A.~Griewank\textsuperscript{2} and  M.~Mueller-Preussker\textsuperscript{3}}
\address{ \textsuperscript{1} NIC, DESY Zeuthen, Platanenallee 6, D-15738 Zeuthen, Germany}
\address{ \textsuperscript{2} Institut f\"ur Mathematik, Humboldt Universit\"at zu Berlin, 
Unter den Linden 6, D-10099 Berlin }
\address{ \textsuperscript{3} Institut f\"ur Physik, Humboldt-Universit\"at zu Berlin, 
Newtonstr. 15, D-12489 Berlin
}
\ead{griewank@mathematik.hu-berlin.de, Karl.Jansen@desy.de, 
leovey@mathematik.hu-berlin.de, mmp@physik.hu-berlin.de, Andreas.Nube@desy.de}

\begin{abstract}
In this project we initiate an investigation of the applicability of 
Quasi-Monte Carlo methods to lattice field theories in order to improve the asymptotic 
error behavior of observables for such theories. In most cases the error of an observable 
calculated by averaging over random observations generated from an ordinary Monte Carlo 
simulation behaves like $N^{-1/2}$, where $N$ is the number of observations. By means 
of Quasi-Monte Carlo methods it is possible to improve this behavior for certain problems 
to up to $N^{-1}$.  We adapted and applied this approach  to simple systems like the 
quantum harmonic and anharmonic oscillator and verified an improved error scaling. 
\end{abstract}

\section{Introduction}
Four-dimensional quantum field theories play a crucial role in the mathematical 
description of the fundamental forces of nature. The path integral formalism developed 
by R.P. Feynman has been established since decades to quantize classical 
field theories and thus, to formulate quantum field theories.\\
A quantum field theory (QFT) describes the behavior of certain particles and their 
interaction. In the absence of an interaction the path integral of a QFT is usually 
trivial. It is mostly the interaction term that makes the path integral challenging 
to evaluate. In cases where the coefficient of the interaction term, the coupling 
constant, is small enough a perturbative treatment of the path integral is possible 
and often sufficient.\\
However many interesting quantities, like e.g. the hadron spectrum, decay constants, 
certain matrix elements and form factors, have to be calculated in a regime of a 
strong coupling(-constant), where a perturbative approach must fail. In such 
situations it is necessary to treat the path integral non-perturbatively. Because of 
the lack of closed form solutions the path integral can only be evaluated numerically. 
In order to do so, it is necessary to give the path integral a mathematically well 
defined meaning. A straightforward method is it to discretize space and time by 
introducing a, moreover Euclidean space-time lattice with a fixed spacing between two 
neighboring lattice points, the lattice spacing.
Restricting the system to a finite lattice extension (and applying certain boundary
conditions) the 
infinite-dimensional path integral is converted to a finite-dimensional integral. 
(In principle, one still has to perform the transition to zero lattice spacing and 
therefore to infinitely many space-time points at the end.)\\
Such lattice path integrals can easily have dimensions of $O(10^9)$ 
(e.g. simulations of lattice-discretized quantum chromodynamics (QCD), the theory of 
strong interactions of elementary particles). 
The high dimensionality of the problem restricts the spectrum of applicable algorithms 
to Monte Carlo-based methods. Especially Markov chain-Monte Carlo (Mc-MC) methods 
have been successfully applied since the beginning of the study of lattice field theories. 
With the algorithms employed observables calculated from a Monte Carlo chain of $N$ 
steps will obey a statistical error proportional to $1/\sqrt{N}$. \\
Recent developments in the field of Quasi-Monte Carlo (QMC) methods, discussed in 
more detail in sections \ref{sec:Plain:Int} to \ref{sec:RQMC}, show that under certain 
conditions it is possible to construct sets of samples of integration points leading 
to much faster rates of convergence of an observable and a much better asymptotic 
error behavior of up to $1/N$.\\
Such an improved error behavior would decrease the number of samples necessary to 
achieve a certain error bound, resulting in a drastic reduction of runtime. Note that 
for present computations in field theory state-of-the-art supercomputers are used.
It is unclear, whether QMC methods can be used for lattice field theory simulations. 
As a first step, to nevertheless investigate this possibility, we will focus in this 
work on the study of much simpler models, namely the quantum mechanical harmonic 
and anharmonic oscillator.

\section{Quantum Mechanical Harmonic and Anharmonic Oscillator}
In this section we will discuss the basic steps for the quantization of the theory 
in the path integral approach and the discretization on a time lattice.
The first step is the construction of the Lagrangian (resp. the action) of the 
corresponding classical mechanical system for a given path $x(t)$ of a particle 
with mass $M_0$. For a numerically stable evaluation of the path integral it is 
essential to pass on to Euclidean time. In this case the Lagrangian $L$ and the 
action $S$ is given by:
\begin{align}
\label{eq:Lagrangian}
  L(x,t) &= \frac{M_0}{2}
  \left(\frac{d x}{dt}\right)^2 + V(x) \\
  S(x) &= \int_0^T \, L(x,t) \; dt  .
\end{align}
Depending on the scenario (harmonic or anharmonic oscillator) the potential $V(x)$ 
consists of two parts 
\begin{equation}
V(x) =  \underbrace{\frac{\mu^2}{2} x^2}_\text{harmonic part} + 
\underbrace{\lambda \, x^4}_\text{anharmonic part} \;,
\end{equation}
such that the parameter $\lambda$ controls the anharmonic part of the theory. 
It should also be mentioned that in the anharmonic case the parameter 
$\mu^2$ can take on negative values, leading then to a double well potential.

The next step is to discretize time into equidistant time slices with a 
spacing of $a$. The path is then only defined on the time slices:
\begin{align}
  t  & \rightarrow t_i = (i-1) \cdot a \quad i =  1 \ldots d \\
 x(t) & \rightarrow x_i = x(t_i) \; .
\end{align}
On the lattice the derivative with respect to the time appearing in 
\eqref{eq:Lagrangian} (first term) will be replaced by the forward finite difference 
$\nabla x_i = \frac{1}{a} ( x_{i+1} - x_i )$. The choice of the lattice derivative 
is not unique and requires special care, particularly if one considers more 
complicated models like lattice QCD. But in \cite{Creutz_and_Freedman} it was 
shown that the lattice derivative chosen here permits a well defined continuum 
limit. Putting all the ingredients together, we can write down the lattice action 
for the (an)harmonic oscillator
\begin{equation}
  S^\text{latt}(x) = a \sum_{i=1}^{ d } \frac{M_0}{2} \left( \nabla x_i \right)^2 + V(x_i) \; .
\end{equation}
For the path a cyclic boundary condition $x_{d+1} = x_1$ can be assumed.
In the following the superscript ``latt'' will be dropped, as we will only 
refer to the lattice action from now on.
The expectation value of an observable $O$ of the quantized theory expressed 
in terms of the path integral reads as follows:
\begin{equation}\label{sec:Plain:OBS}
\left\langle O(x) \right\rangle \, = \, \frac{\int_{\mathbb{R}^d} O(x)
e^{-S(x)} d x_1...d x_d }{\int_{\mathbb{R}^d} e^{-S(x)} d x_1...d x_d }\;.
\end{equation}
This expression is suitable for a numerical evaluation of certain quantities of 
the underlying theory. Up to now only Monte Carlo methods are known to give 
reliable results for dimensions $d \gg 10$. One type of such methods, often 
used in physics, is the Markov chain-Monte Carlo approach mostly applying 
the weight $\propto e^{-S(x)}$ for sampling paths $\{x_i\}$ (so-called 
``importance sampling''). Especially the Metropolis algorithm\cite{Metropolis} 
is suitable and a straightforward solution of \eqref{sec:Plain:OBS}
(also described in \cite{Creutz_and_Freedman}) and serves as a reference 
method for the QMC approach, which is much less intuitive.
The theory of QMC methods is a purely mathematical topic.
During the discussion of the key aspects of QMC, following in the next 
sections, we will stick to a rather mathematical language, being more 
adequate for the description of a mathematical issue.

\section{Direct Monte Carlo and quasi--Monte Carlo methods}\label{sec:Plain:Int}
In many practical applications one is interested in calculating quotients of the form 
\eqref{sec:Plain:OBS} where the action $S(.)$ and the observables $O(.)$ are 
usually smooth functions in high dimensions. In some special situations 
where one would like to deal with integrands of moderately high dimensions, one may consider an estimator for the integral 
$I_1$ in the numerator and $I_2$ in the denominator of \eqref{sec:Plain:OBS} separately, 
and then take $I_1/I_2$ as an estimation of $\left\langle O(x) \right\rangle$. 
Another possibility is to take a joint estimator for the total quantity 
$\left\langle O(x) \right\rangle$ using a single direct sampling method. 
A well known approach based on direct sampling is the so called weighted 
uniform sampling (WUS) estimator, analyzed in \cite{PowellSwann66}.
We will show some characteristics of the WUS estimator in section \ref{sec:WUS}, and we will 
refer from now on to these methods as \textit{plain} or \textit{direct} sampling methods 
for estimating \eqref{sec:Plain:OBS}.
In many interesting examples, we encounter the case were the action $S(.)$ 
and the observable $O(.)$ lead to integrals $I_1$,$I_2$ of Gaussian type. 
Then the integrals $I_1$,$I_2$ can be written in the form 
\[
I_i\:=\: \frac{1}{(2\pi)^{d/2} \sqrt{\det(C)}} 
        \int_{\mathbb{R}^d}g_i(\vx)e^{-\frac{1}{2} \vx^\top C^{-1} \vx} d\vx, 
	\quad
	\vx=(x_1,\dots,x_d), \; i=1,2 \quad ,
\]
where $C$ denotes the covariance matrix of the Gaussian density function. 
A transformation to the unit cube in $\mathbb{R}^d$ can be applied such 
that the corresponding integrals take the form 
\begin{equation}\label{gen:expected_v}
I
\:=\:  \int_{[0,1]^d}g(A \vPhi^{-1}(\vz))d\vz
\:=\:  \int_{[0,1]^d}f(\vz)d\vz
\:=\: I_{[0,1]^d}(f)
	,\quad
	\vz=(z_1,\dots,z_d)\,.
\end{equation}
Here $AA^\top=C$ is some symmetric factorization of the covariance matrix, 
and $\vPhi^{-1}(\vz):=(\Phi^{-1}(z_1),\dots,\Phi^{-1}(z_d))^\top$, 
where $\Phi^{-1}({\cdot})$ represents 
the inverse of the normal cumulative distribution function $\Phi({\cdot})$.\\
In the classical plain or direct Monte--Carlo (MC) approach one tries
to estimate \eqref{gen:expected_v} 
by generating samples pseudo-randomly. One starts with a finite sequence of 
independent identically distributed (i.i.d.) samples $P_N=\{\vz_1,\dots,\vz_N\}$, 
where the points $\vz_j, \; 1\le j \le N$, 
have been generated from the uniform distribution in $[0,1]^d$. 
Then, the quadrature rule is fixed by taking the average of the function evaluations for $f$ 
\[
Q_N:= \frac{1}{N} \sum_{j=1}^{N} f(\vz_j),
\]   
as an approximation of the desired integral $\int_{[0,1]^d} f(\vz) \; d \vz$. 
The resulting estimator $\hat{Q}_N$ is unbiased. The integration error 
can be approximated via the central limit theorem, given that $f$ belongs to $L_2([0,1]^d)$. 
The variance of the estimator $\hat{Q}_N$ is given by
$$
\frac{\sigma^2}{N}=\frac{1}{N}\left( \int_{[0,1]^d} f^2(\vz) \; d\vz
  - \left(\int_{[0,1]^d} f(\vz) \; d\vz \right)^2 \right).
$$    
As measured by its standard deviation from zero  
the integration error associated with the 
MC approach is then of order $O(N^{-\frac{1}{2}})$.
The quality of the MC samples relies on the selected pseudo--random 
number generators of uniform samples, here we use the \textit{Mersenne Twister} generator from Matsumoto and Nishimura (see \cite{Matsumoto98}).
MC is in general a very reliable tool in high--dimensional integration, 
but the order of convergence is in fact rather poor. 

In contrast, quasi--Monte Carlo (QMC) methods generates deterministically 
point sets that are more regularly distributed than 
the pseudo--random points from MC (see \cite{L'Ecuyer01}, \cite{Novak_and_Wozniakowski2}, 
\cite{DiPi10}, \cite{KSS_Review12}).  
Typical examples of QMC are shifted 
lattice rules and low--discrepancy sequences.   
To explain what we mean by ``regularly distributed'', 
we define now the classical notion of discrepancy 
of a finite sequence of points $P_N$ in $[0,1)^d$.    
Given $P_N=\{\vz_{1},\dots,
\vz_{N}\}$ a set of points in $ [0,1)^{d}$, 
and a nonempty family $\mathbb{I}$ of Lebesgue-measurable 
sets in $[0,1)^{d}$, we define the classical discrepancy function by
\[D(\mathbb{I};P_N) := \sup_{B \in \mathbb{I}}\left|\frac{\sum_{i=1}^{N}\:
c_{B}(\vz_{i})}{N}-\lambda_{d}(B)\right|,\]
where $c_{B}$ is the characteristic function of $B$.
This allows us to define the so called \textit{star discrepancy}.

\begin{definition}
We define the \textit{star discrepancy} $D^{\star}(P_N)$ of the point set $P_N$ 
by $D^{\star}(P_N):=D (\mathbb{I};P_N)$, where $\mathbb{I}$ is the family of all 
sub-intervals of the form $\prod_{i=1}^{d}[0,u_{i})$,
with $u_{i}\ge 0, \; 1\le i \le d$.
\end{definition}
The \textit{star discrepancy} can be considered as a measure of the worst difference 
between the uniform distribution and the sampled distribution in $ [0,1)^{d}$ 
attributed to the point set $P_N$. 
The usual way to analyze QMC as a deterministic method 
is by choosing a class of integrand functions $F$, and a 
measure of discrepancy $D(P_N)$ for the point sets $P_N$.
Then, the deterministic integration error is usually given in the form
$$
|Q_N -\int_{[0,1]^d} f(\vz) \; d \vz | \;\; \le \; D(P_N) V(f), 
$$
where $V(f)$ measures a particular variation of the 
function $f \in F$. A classical particular error bound in this form is 
the famous Koksma--Hlawka inequality, 
where $D(P_N)$ is taken to be the \textit{star discrepancy} 
of the point set $P_N$, and $V(f)$ is the variation 
in the sense of Hardy and Krause of $f$. 
In the context of QMC, a sequence of points  in $ [0,1)^{d}$ 
is called a low--discrepancy sequence 
if $D^{\star}(P_N)=O(N^{-1}(\log(N))^{d})$ 
for all truncations of the sequence to its first $N$ terms. \\

\subsection{Quasi--Monte Carlo errors and complexity}
There are certain reproducing kernel Hilbert spaces $\mathbb{F}_{d}$ of functions
$f:[0,1]^{d}\to\mathbb{R}$ which are particularly useful for estimating the
quadrature error of QMC methods (see \cite{Hick98}). Consider a kernel 
$K:[0,1]^{d}\times[0,1]^{d}\to\mathbb{R}$ satisfying 
$K(\cdot,\vy)\in\mathbb{F}_{d}$ and $\langle f,K(\cdot,\vy)\rangle=f(\vy)$ for
each $\vy\in[0,1]^{d}$ and $f\in\mathbb{F}_{d}$. We denote now with 
$\langle\cdot,\cdot\rangle$ and $\|\cdot\|$ the inner product
and norm in $\mathbb{F}_{d}$. If the integral
$$
I(f)=\int_{[0,1]^{d}}f(\vz)d\vz
$$
is a continuous functional on the space $\mathbb{F}_{d}$, 
then the worst case quadrature error $e_{N}(\mathbb{F}_{d})$ 
for point sets  $P_N=\{\vz_{1},\dots,\vz_{N}\}$ and quasi-Monte Carlo algorithms
for the space $\mathbb{F}_{d}$ can be given by 
\[
e_{N}(\mathbb{F}_{d}):=\sup_{f\in\mathbb{F}_{d}\,,\|f\|\le 1}|
I(f)-Q_{N}(f)|=\sup_{\|f\|\le 1}|\langle
f,h_{N}\rangle|=\|h_{N}\|,
\]
due to Riesz' representation theorem for linear bounded functionals. In this case, the {\em
representer} $h_{N}\in\mathbb{F}_{d}$ of the quadrature error is given by 
$$
h_{N}(\vz)=\int_{[0,1]^{d}}K(\vz,\vy)d\vy -
\frac{1}{N}\sum_{i=1}^{N}K(\vz,\vz_{i})\quad(\forall
\vz\in[0,1]^{d}).
$$
In QMC error analysis, one usually considers the weighted (anchored) 
tensor product Sobolev space introduced in 
\cite{SlWo98}
\[
\mathbb{F}_{d}=\mathcal{W}_{2,{\rm mix}}^{(1,\ldots,1)}([0,1]^{d})
=\bigotimes_{i=1}^{d}W_{2}^{1}([0,1]) \; ,
\]
with the weighted norm $\|f\|_{\gamma}^{2}=\langle
f,f\rangle_{\gamma}$ and inner product 
\[
\langle f,g\rangle_{\gamma}=\sum_{u\subseteq\{1,\ldots,d\}}
\prod_{j\in u}\gamma_{j}^{-1}
\int_{[0,1]^{|u|}}\frac{\partial^{|u|}}{\partial
\vz_{u}}f(\vz_{u},\mathbf{1})\frac{\partial^{|u|}}{\partial
\vz_{u}}g(\vz_{u},\mathbf{1})d \vz_{u},
\]
where for $u \subseteq \{1,\dots,d\}$ we denote by $|u|$ its cardinality, 
and $(\vz_{u},\mathbf{1})$ denotes the vector 
containing the coordinates of $\vz$ with indices in $u$, and the other 
coordinates set equal to $1$. 

The corresponding reproducing kernel is given by 
\[
K_{d,\gamma}(\vz,\vy)=\prod_{j=1}^{d}(1+\gamma_{j}
[1-\max(z_{j},y_{j})])\quad(\vz,\vy \in[0,1]^{d}).
\]
There are several other examples considered for error analysis. For example, the 
weighted Walsh space consisting of Walsh series
(see \cite[Example 2.8]{DiPi10} and \cite{Dick08}). 
The weighted tensor product Sobolev space allow for explicit QMC constructions 
deriving error estimates of the form
\begin{equation}\label{rate}
e_{N}(\mathbb{F}_{d})\leq C(\delta)N^{-1+\delta}
\quad(\delta\in(0,\textstyle{\frac{1}{2}}]),
\end{equation}
where the constant $C(\delta)$ is independent on the dimension $d$,
given that the sequence of weights $(\gamma_{j})$ satisfies (see \cite{Kuo2003})
\[
\sum_{j=1}^{\infty}\gamma_{j}^{\frac{1}{2(1-\delta)}}<\infty\,.
\]
Traditional unweighted function spaces considered for integration 
suffer the from the curse of dimensionality. Their weighted variants describe a 
setting where the variables or group of variables may vary in importance. 
Thus, they give a partial explanation of why some 
very high-dimensional spaces become tractable for QMC.


Explicit QMC constructions satisfying \eqref{rate} are \textit{shifted lattice rules} 
for weighted spaces.  
The rate (\ref{rate}) can be also obtained for Niederreiter and Sobol' sequences (see \cite{Wang03}).

The idea of ``weighting'' the norm of the spaces to obtain tractable results can be applied in fact to  
more general function spaces than smooth function spaces of tensor product form, and many integration 
examples can be found in \cite{Novak_and_Wozniakowski2}. 
In our numerical experiments, we used so far QMC algorithms based on 
a particular type of low--discrepancy sequences. 
Numerical experiments with shifted lattice rules will be carried out in the near future, following 
new techniques for fixing adequate weights introduced in \cite{GLLZ12}.

\section{Low--discrepancy $(t,d)$-sequences}

The most well known type of low--discrepancy sequences are the so called
$(t,d)$-sequences. 
To introduce how $(t,m,d)$-nets and $(t,d)$-sequences are defined, 
we consider first \textit{elementary intervals} in a integer base $b \ge 2$.
Let $E$ be any sub-interval of $[0,1)^{d}$ of the form 
$E=\prod_{i=1}^{d}[a_{i}b^{-c_{i}},(a_{i}+1)b^{-c_{i}})$
with  $a_{i},\: c_{i} \: \in \mathbb{N} ,  c_{i} \ge 0, \:0\leq a_{i} < b^{-c_{i}}$ 
for $1\leq i\leq d$. An interval of this form is 
called an elementary interval in base $b$.
\begin{definition}
 Let $\:0\leq t \leq m$ be integers. A $(t,m,d)$-net in base $b$ is a point
 set $P_N$ of $N=b^{m}$ points in  $[0,1)^{d}$ 
such that  every elementary interval 
$E$ in base $b$ with $\lambda_{d}(E)=\frac{b^{t}}{b^{m}}$ 
contains exactly $b^{t}$ points.
\end{definition}

\begin{definition}
 Let $t\geq 0$ be an integer. A sequence $\mathbf{x}_{1},\mathbf{x}_{2},...$ 
of points in  $[0,1)^{d}$ is a $(t,d)$-sequence in base $b$ if for all integers 
$k\geq0$ and $m>t$, the point set consisting of $N=b^{m}$ points 
$\mathbf{x}_{i}$ with $kb^{m}\leq i < (k+1)b^{m}$,   
is a $(t,m,d)$-net in base $b$.
\end{definition}
The parameter $t$ is called the \textit{quality parameter} 
of the $(t,d)$--sequences.
In \cite{Nied92}, theorem 4.17, it is shown that $(t,d)$-sequences 
are in fact low--discrepancy sequences. We reproduce this result in the following
\begin{theorem}
The star-discrepancy $D^{\star}$ of the first $N$ terms $P_N$ of a $(t,d)$-sequence in 
base $b$, satisfies
$$
 N D^{\star}(P_N) \leq C(d,b) b^t (log(N))^d + O(b^t (log(N))^{d-1}),
$$
where the implied constants depend only on $b$ and $d$.
If either $d=2$ or $b=2$, $d=3,4$, we have
$$
C(d,b)=\frac{1}{d}\left( \frac{b-1}{2 log(b)} \right)^d,
$$
and otherwise
$$
C(d,b)=\frac{1}{d!} \frac{b-1}{2 \lfloor b/2 \rfloor} \left( \frac{\lfloor b/2 \rfloor}{log(b)} \right)^d. 
$$
\end{theorem}

Explicit constructions of $(t,d)$-sequences are available. Some of them are 
the generalized Faure, Sobol', Niederreiter and Niederreiter--Xing sequences. 
All these examples fall into the category of constructions 
called \textit{digital sequences}. 
We refer to \cite{DiPi10} for further reading on this topic. 

\section{Randomized QMC}\label{sec:RQMC}

There are some advantages in retaining the probabilistic properties of the sampling.  
There are practical hybrid methods permitting us to combine the good features of MC and 
QMC. Randomization is an important tool for QMC if we are interested for a practical
error estimate of our sample quadrature $Q_N$ to the desired integral. One goal is 
to randomize the deterministic point set $P_N$ generated by QMC in a way that 
the estimator $\hat{Q}_N$ preserves unbiasedness. Another important goal is to preserve 
the better equidistribution properties of the deterministic construction. 
 
The simplest form of randomization applied to \textit{digital sequences} seems to be 
the technique called \textit{digital $b$--ary shifting}. In this case, we add 
a random shift $\Delta \in [0,1)^d$ to each point of the deterministic set 
$P_N=\{\vz_{1},...,\vz_{N}\}$ using 
operations over the selected ring $\mathbb{F}_b$. 
The application of this randomization preserves in particular the $t$ value of any projection of 
the point set (see \cite{L'Ecuyer01} and references therein). The resulting estimator is 
unbiased.\\
The second randomization method we present is the one introduced by  
Art B. Owen (\cite{OWE95}) in 1995. He considered $(t,m,d)$-nets and $(t,d)$-sequences 
in base $b$ and applied a randomization procedure based on permutations of the digits of 
the values of the coordinates of points in these nets and sequences. This can be interpreted 
as a random scrambling of the points of the given sequence in such a way 
that the net structure remains unaffected.
We do not discuss here in detail Owen's randomization procedure, 
or from now on called \textit{Owen's scrambling}. 
The main results of this randomization procedure can be stated in the following
\begin{proposition}(\textbf{Equidistribution})\\
A randomized $(t,m,d)$-net in base $b$ using Owen's scrambling is again a $(t,m,d)$-net 
in base $b$ with probability 1. A randomized $(t,d)$-sequence in base $b$ using Owen's 
scrambling is again a $(t,d)$-sequence in base $b$ with probability 1.
\end{proposition}
\begin{proposition}(\textbf{Uniformity})\\
Let $\tilde{\vz}_i$ be the randomized version of a point 
$\vz_i$ originally belonging  to a  $(t,m,d)$-net 
in base $b$ or a $(t,d)$-sequence in base $b$, using Owen's scrambling. 
Then $\tilde{\vz}_i$ has 
the uniform distribution in $[0,1)^d$, that is, for any Lebesgue measurable set $G \subseteq 
[0,1)^d$ , $P( \tilde{\vz}_i \in G)= \lambda_d(G)$, 
with $\lambda_d$ the $d$-dimensional Lebesgue measure.  
\end{proposition} 

The last two propositions state that after \textit{Owen's scrambling} of \textit{digital sequences} 
we retain unaffected the low discrepancy properties of the constructions, and that 
after this randomization procedure we obtain random samples uniformly distributed in $[0,1)^s$. \\

The basic results about the variance of the randomized QMC estimator $\hat{Q}_N$ 
after applying \textit{Owen's scrambling} 
to $(t,m,d)$-nets in base $b$ (or of $(t,d)$-sequences in base $b$ ) 
can be found in \cite{Owen97}. We summarize these results in the following

\begin{theorem}
Let $\tilde{\vz}_i$, $1\le i \le N$, be the points of a 
scrambled $(t,m,d)$-net in base $b$, and let $f$ be a function 
on $[0,1)^d$ with integral $I$ and variance $\sigma^2=\int (f-I)^2 d\vz  < \infty.$ 
Let $\hat{Q}_N= N^{-1} 
\sum_{i=1}^N f(\tilde{\vz}_i)$, where $N=b^m$. 
Then for the variance $V(\hat{Q}_N)$ of the randomized QMC estimator 
it holds
\[ V(\hat{Q}_N)=o(1/N), \: \text{ as } N \rightarrow \infty,  \quad \text{and} \quad 
 V(\hat{Q}_N)\leq \frac{b^t}{N}\left( \frac{b+1}{b-1} \right)^d \sigma^2.\]
For $t=0$ we have
\[ V(\hat{Q}_N)\leq \frac{1}{N}\left( \frac{b}{b-1} \right)^{d-1} \sigma^2.\]
\end{theorem}
The above theorem says that the variance of scrambled $(0,m,d)$--nets is never more than 
$3$ times the variance of the corresponding MC estimator.
The bound of the theorem above can be improved (see theorem 13.9 in \cite{DiPi10}) to show that the 
variance of scrambled $(0,m,d)$--nets are in fact always smaller than the variance of the MC estimator.
If the integrand at hand is smooth enough, using \textit{Owen's scrambling} 
it can be shown that one can obtain an improved asymptotic error estimate of order
$O(N^{-\frac{3}{2}-\frac{1}{d}+\delta})$, for any $\delta>0 $, see \cite{Owen08}. 
Improved scrambling techniques have been developed in \cite{MAT98},\cite{Tezuka03}.\\

\section{Weighted uniform sampling}\label{sec:WUS}
Weighted uniform sampling is a way of estimating a quotient of integrals of the form 
\[
R:=\frac{ \int_{[0,1]^d}f_1(\vz)d\vz}{ \int_{[0,1]^d}f_2(\vz)d\vz}
\] 
by taking the estimator 
\begin{equation}
\label{eq:WUS}
\hat{R}_N:=\frac{ \sum_{j=1}^N f_1(\vz_j)}{\sum_{j=1}^N f_2(\vz_j)} \; ,
\end{equation}
where the points $\vz_j, \; 1\le j \le N$, 
have been generated from the uniform distribution in $[0,1]^d$. 
This estimator was analyzed in \cite{PowellSwann66} and applications have been 
investigated for example in \cite{SpaMa} and \cite{Caflisch95}. The bias and the root mean 
square error (RMSE) of this estimator satisfy 
\begin{align*}
& Bias(\hat{R}_N)=\frac{R \,var(f_2)}{N} -\frac{cov(f_1,f_2)}{N} + O(N^{-\frac{3}{2}}) \\
& RMSE(\hat{R}_N)=\frac{\sqrt{var(f_1) + R^2 var(f_2) - 2R\, cov(f_1,f_2)}}
{\sqrt{N}} + O(N^{-\frac{3}{4}}) \; .
\end{align*}
The bias of the estimator is asymptotically negligible compared with the RMSE.
One clear disadvantage of WUS against Mc-MC or Importance Sampling 
for problems with large regions of relative low values 
of the integrands is that with WUS we sample over 
the entire unit cube $[0,1]^d$ uniformly, 
while Mc-MC and Importance Sampling based techniques try to concentrate 
in more characteristic or important regions of the integrands.   
These limitations where observed in our numerical experiments.

\section{Numerical experiments}
We consider for our numerical tests the \textit{quantum mechanical 
harmonic and anharmonic oscillator} in the \textit{path integral
approach}  as described in section 2.
For definiteness we repeat here the expression for the action of the system:
 \begin{equation}
\label{eq:action_detail}
 S(x)=\frac{a}{2} \sum_{i=1}^d \frac{M_0}{a^2} (x_{i+1}-x_i)^2 + \mu^2 x_i^2 
 + 2 \lambda x_i^4 \; .
 \end{equation}
We investigate the two observable functions
\[
O_1(x)=\frac{1}{d}\sum_{i=1}^d x_i^2 \, , \;
O_2(x)=\frac{1}{d}\sum_{i=1}^d x_i^4 \; ,
\]
using the notation $\left\langle X^2 \right\rangle$,$\left\langle X^4 \right\rangle$ 
for $\left\langle O_1(x) \right\rangle$,$\left\langle O_2(x) \right\rangle$ in our tests. 
\subsection{Harmonic Oscillator}
\label{ssec:HO}
\begin{figure}[ht]
\centering
  \begin{minipage}[b]{0.8\linewidth}
    \centering
  \includegraphics[width=\textwidth]{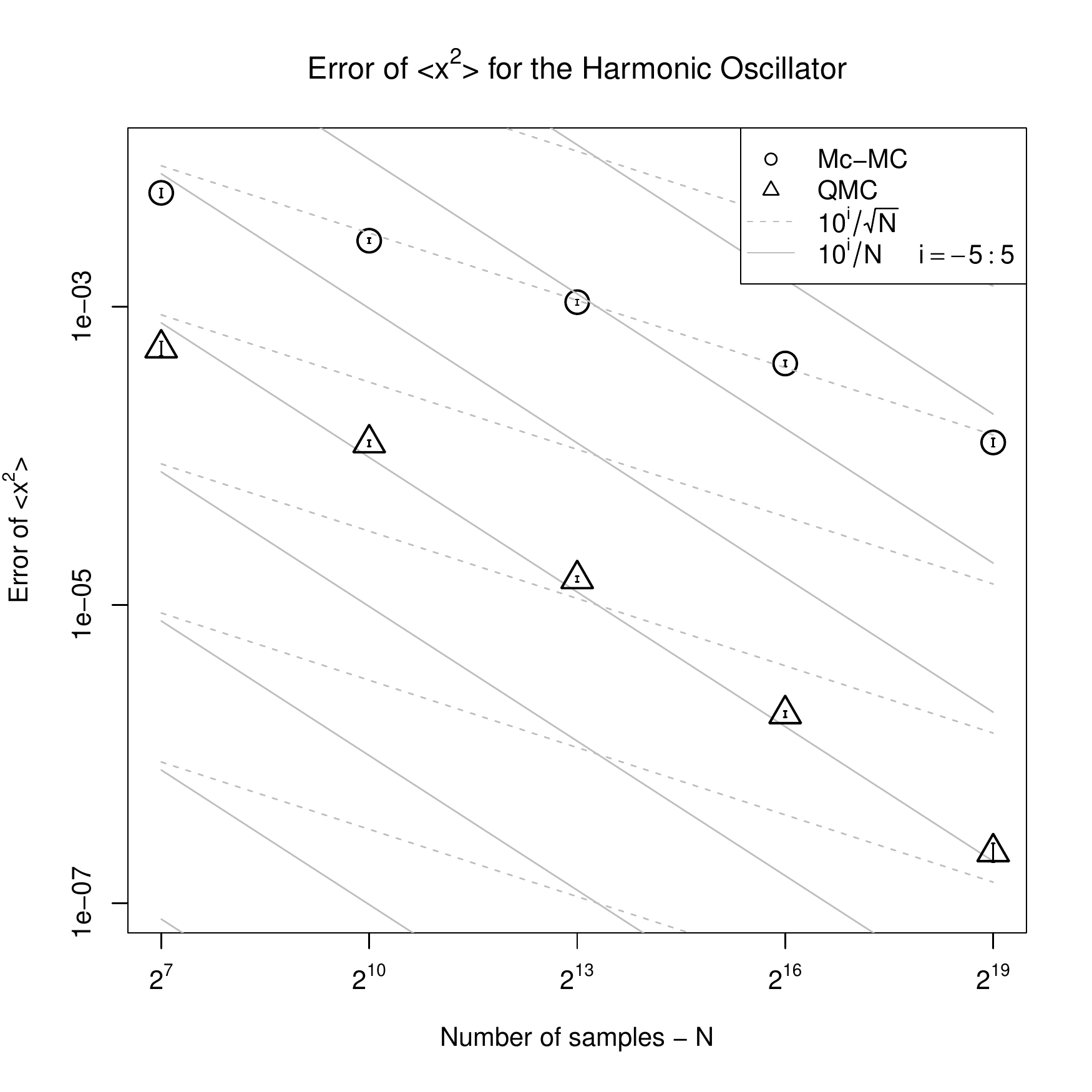}
  \caption{error of $\langle X^2 \rangle$ in dependence of the number of samples $N$, $\lambda=0$ (harmonic oscillator),$d=51$, $M_0=0.5$ and $\mu^2=2.0$}
  \label{fig:x2_harmonic}
\end{minipage}
\end{figure}

For the harmonic oscillator we can apply immediately the direct sampling approach described in sections \ref{sec:Plain:Int} and \ref{sec:WUS} for calculating estimates of observables $O$ by setting
\[
f_1 = O( A \Phi^{-1}(\vz) ) \; , \;\; f_2 = 1
\]
in \eqref{eq:WUS}.
The matrix $A$ is a square root of $C$, the covariance matrix of the variables $x_i$, appearing in the action if it is expressed as a bilinear form: $S=\frac{1}{2}x^T C^{-1} x$.
Different factorizations, namely Cholesky and PCA (principle component analysis) have been tried out. The PCA based factorization seemed to perform better in our tests, which is the reason why we will only show results for this method.
The PCA can be explicitely obtained for circulant Toeplitz matrices and the matrix--vector products can be efficiently computed by means of the fast Fourier transform.
In the ordinary Mc-MC approximation, we used the Mersenne Twister\cite{Matsumoto98} pseudo random number generator.
For the QMC tests, we use randomly scrambled Sobol' sequences using the technique proposed by J. Matous\v{e}k\cite{MAT98}.
The error of $\langle X^2 \rangle$ was obtained by scrambling 10 times the QMC sequence and making 10 runs of an Mc-MC simulation (with different seeds). This procedure is repeated 30 times in both cases to obtain the error of the error.
From the results,  shown in figure \ref{fig:x2_harmonic}, we can see a scaling that agrees perfectly with the expected behavior, namely $N^{-0.5}$ for Mc-MC and $N^{-1}$ for QMC, for large $N$.
Although this example is trivial, it was our first successful application of the QMC approach in a physical model and motivated us to pass on to more complicated models.


\subsection{Anharmonic Oscillator}
The WUS approach was also used for this problem to estimate $\langle X^4 \rangle$ and $\langle X^2 \rangle$.

\begin{figure}
\centering
\begin{subfigure}[b]{0.8\textwidth}
\includegraphics[width =\textwidth]{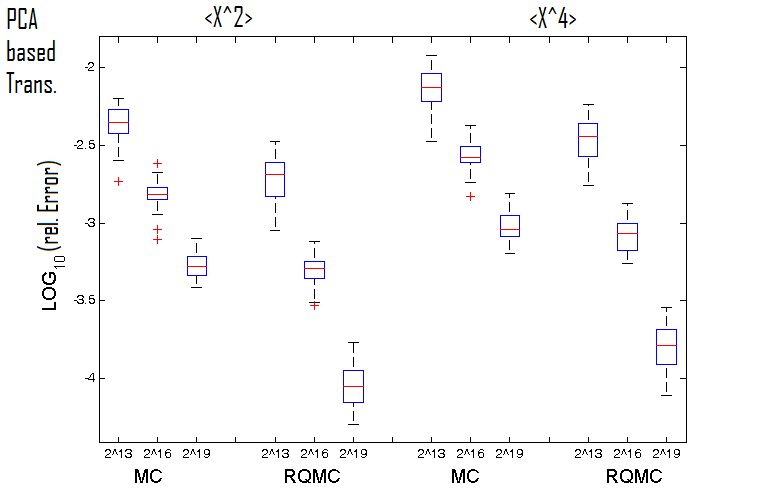}
\caption{$a=0.015$, $d=100$ ($T=1.5$)}
\end{subfigure}
\begin{subfigure}[b]{0.8\textwidth}
\includegraphics[width =\textwidth]{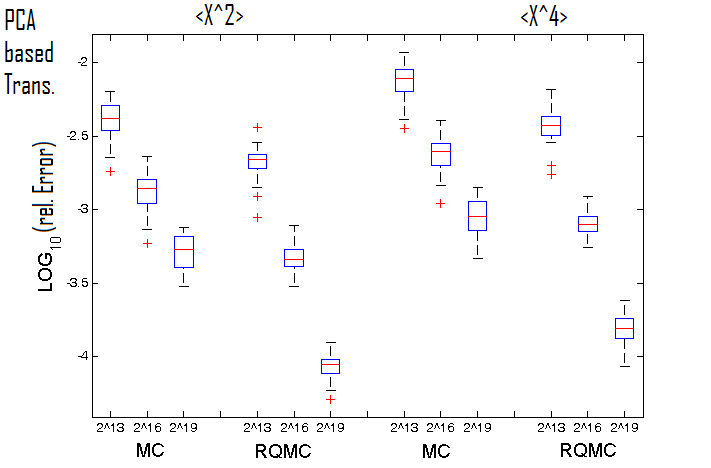}
\caption{$a=0.0015$, $d=1000$ ($T=1.5$)}
\end{subfigure}
\caption{Shown is the $log_{10}$(relative error) as box plots with 30 repetitions of the experiment with $\lambda=1.0$, $\mu^2=-16$ and $a$ and $d$ as indicated.
  For the sample generation MC and randomly scrambled Sobol' (RQMC) was used with $2^{13},2^{16}$ and $2^{19}$ points,
  . The approximate convergence rate is of $O(N^{-0.75})$ for RQMC.
}
\label{sec:numex:fig:1}
\end{figure}

 With the anharmonic term in the action the distribution function of the variables $x_i$ becomes very complicated. This makes it very hard to generate the samples directly from the PDF of the anharmonic oscillator. Instead of this, the anharmonic term and a part of the harmonic term is treated as part of the weight functions $f_1$ and $f_2$ in \eqref{eq:WUS},
leaving the sampling procedure of the $x_i$ as it was for the harmonic oscillator, accept for a different factor $\mu^2_{sim}$ in front of the harmonic term\begin{equation}
\label{eq:weight_fns_anharm}
f_1(\vz) = O( A \Phi^{-1}(\vz) ) f_2(\vz) \; ,\quad 
f_2(\vz) = e^{ - \sum a\left( \frac{\mu^2-\mu_{sim}^2}{2}\right) (A \Phi^{-1}(\vz))_i^2 +  a \lambda (A \Phi^{-1}(\vz))_i^4 }\; .
\end{equation}
This procedure is neccessary, because of $C=A^T A$ being positive definite only if $\mu^2_{sim} > 0 $, which is neccessary for the existence of $A^{-1}$ during the sampling procedure.
Further, it is important to note that the PCA factorization during the generation of the gaussian samples is essential for an efficient reduction of the effective dimension (see \cite{CAF97}) of the problem. For the parameters listed below, we estimated the effective dimensions of the functions \ref{eq:weight_fns_anharm} to be close to $20$ (for a $99 \%$ variance concentration). On the other hand we found out that the effective dimension depends also very strong on the parameter $T = d a$, the physical time extent of the system. For small $T$-values, say $T < 0.2$, the effective dimension is reduced sufficiently good like in the harmonic case, such that the QMC approach leads to a $1/N$ error scaling. The situation changes for $T=1.5$, where the error behaves only like $1/N^{0.75}$, due to the increase of the effective dimension.
The parameters were set to $M_0 = 0.5$, $\lambda=1.0$, $\mu^2 = -16$.
In the two tests the parameters $a$ and $\mu^2_{sim}$ had been adjusted such, that $ T $ was kept fixed. We set $a=0.015$ and $\mu^2_{sim} = 0.015$ for $d=100$, whereas  for $d=1000$ $a=0.0015$ and $\mu^2_{sim}=0.0015$ was chosen.
The error analysis of $\langle X^2 \rangle$ and $\langle X^4 \rangle$ has been adopted from the harmonic oscillator test case described in the last subsection \ref{ssec:HO}. The result is shown in figure \ref{sec:numex:fig:1}.
For reasons mentioned earlier, WUS shows its limitations for large $T$ in our experiments. 
If $T\geq5$ and $\mu^2 \leq -4 $, then we observe poor results with the Mc-MC or RQMC direct WUS 
sampling method. For $T \in [1,1.5]$ and $\mu^2 \in [-20,10] $ the PCA  results for RQMC seem satisfactory. The resulting 
estimation of the ground state energy matches in at least two significant digits with the theoretical 
value, $E_0 = 3.863...$, calculated in \cite{Blank79}, namely $\hat{E}_0 = 3.856 \pm 0.004 $ for 
$d=100$ and $\hat{E}_0 = 3.864\pm0.003$ for $d=1000$.

\section{Concluding Remarks}
For the harmonic oscillator we found a large-$N$ error behavior as expected 
for QMC ($\sim 1/N $) and Mc-MC ($\sim 1/\sqrt{N}$).
Also for the anharmonic oscillator the estimation procedure leads to a significant 
improvement when employing the  QMC approach. In this case, the error scaling is 
only of $O(N^{-0.75})$ instead of the theoretically best case of $O(N^{-1})$.
Further, we found that the applicability of the WUS approach seems to be limited by the 
physical time extent $T=d a$. Stable results could only by found for values $T\leq 1.5$. 
On the other hand, the choice of $a$ does not seem to have any effect and the 
accessible range of $T$ values gives already estimates of the 
ground state energy, compatible (within errors) with the theoretical prediction 
(valid in the limit $T\rightarrow \infty$ and $a\rightarrow 0$).
For the case that the improved error scaling and the mild dependence on the lattice spacing $a$ found here will also be present in more elaborate models, the QMC has the potential to become very valuable in the future.

\section*{Acknowledgement}
The authors wish to express their gratitude to Alan Genz (Washington State University) 
and Frances Kuo (University of New South Wales, Sydney) for inspiring comments and 
conversations, which helped to develop the work in this report. Frances Kuo 
collaborated with us during her visit to the Humboldt-University Berlin in 2011. 
A.N., K.J. and M.M.-P. acknowledge financial support by the DFG-funded corroborative 
research center SFB/TR9. 

\section*{References}

\bibliographystyle{iopart-num}
\bibliography{Articles}

\providecommand{\newblock}{}
\begin{thebibliography}{10}
\expandafter\ifx\csname url\endcsname\relax
  \def\url#1{{\tt #1}}\fi
\expandafter\ifx\csname urlprefix\endcsname\relax\def\urlprefix{URL }\fi
\providecommand{\eprint}[2][]{\url{#2}}

\bibitem{Creutz_and_Freedman}
Creutz M and Freedman B~A 1981 {\em Ann. Phys.\/} {\bf 132} 427--462

\bibitem{Metropolis}
Metropolis N, Rosenbluth A, Rosenbluth M, Teller A and Teller E 1953 {\em J.
  Chem. Phys.\/} {\bf 21} 1087--1092

\bibitem{PowellSwann66}
Powell M~J~D and Swann J 1966 {\em J.Inst.Maths Applics\/} {\bf 2} 228--236

\bibitem{Matsumoto98}
Matsumoto M and Nishimura T 1998 {\em ACM Trans. Model. Comput. Simul.\/} {\bf
  8} 3--30 ISSN 1049-3301
  \urlprefix\url{http://doi.acm.org/10.1145/272991.272995}

\bibitem{L'Ecuyer01}
L'Ecuyer P and Lemieux C 2005 {\em Modeling Uncertainty\/} ({\em International
  Series in Operations Research \& Management Science\/} vol~46) ed Dror M,
  L’Ecuyer P and Szidarovszky F (Springer US) pp 419--474 ISBN
  978-0-7923-7463-3 \urlprefix\url{http://dx.doi.org/10.1007/0-306-48102-2_20}

\bibitem{Novak_and_Wozniakowski2}
Novak E and Wo{\'z}niakowski H 2010 {\em Tractability of multivariate problems.
  {V}olume {II}: {S}tandard information for functionals\/} ({\em EMS Tracts in
  Mathematics\/} vol~12) (European Mathematical Society (EMS), Z\"urich) ISBN
  978-3-03719-084-5 \urlprefix\url{http://dx.doi.org/10.4171/084}

\bibitem{DiPi10}
Dick J and Pillichshammer F 2010 {\em Digital Nets and Sequences: Discrepancy
  Theory and Quasi-Monte Carlo Integration\/} (New York, NY, USA: Cambridge
  University Press) ISBN 0521191599, 9780521191593

\bibitem{KSS_Review12}
Kuo F, Schwab C and Sloan I 2012 {\em ANZIAM Journal\/} {\bf 53}

\bibitem{Hick98}
Hickernell F~J 1998 {\em Math. Comp\/} {\bf 67} 299--322

\bibitem{SlWo98}
Sloan I~H and Wozniakowski H 1997 {\em J. Complexity\/} {\bf 14} 1--33

\bibitem{Dick08}
Dick J 2008 {\em SIAM J. Numer. Anal.\/} {\bf 46} 1519--1553 ISSN 0036-1429
  \urlprefix\url{http://dx.doi.org/10.1137/060666639}

\bibitem{Kuo2003}
Kuo F~Y 2003 {\em J. Complexity\/} {\bf 19} 301--320 ISSN 0885-064X numerical
  integration and its complexity (Oberwolfach, 2001)
  \urlprefix\url{http://dx.doi.org/10.1016/S0885-064X(03)00006-2}

\bibitem{Wang03}
Wang X 2003 {\em Math. Comput.\/} {\bf 72} 823--838 ISSN 0025-5718
  \urlprefix\url{http://dx.doi.org/10.1090/S0025-5718-02-01440-0}

\bibitem{GLLZ12}
Griewank A, Lehmann L, Leovey H and Zilberman M 2012 {\em Math. Comp.\/}

\bibitem{Nied92}
Niederreiter H 1992 {\em Random number generation and quasi-{M}onte {C}arlo
  methods\/} ({\em CBMS-NSF Regional Conference Series in Applied
  Mathematics\/} vol~63) (Philadelphia, PA: Society for Industrial and Applied
  Mathematics (SIAM)) ISBN 0-89871-295-5

\bibitem{OWE95}
Owen A~B 1995 {\em {M}onte {C}arlo and Quasi-{M}onte {C}arlo Methods in
  Scientific Computing\/} ({\em Lecture Notes in Statistics\/} vol 106) ed
  Niederreiter H and Shiue P~J~S (Springer-Verlag) pp 299--317

\bibitem{Owen97}
Owen A~B 1997 {\em SIAM J. Numer. Anal.\/} {\bf 34} 1884--1910 ISSN 0036-1429
  \urlprefix\url{http://dx.doi.org/10.1137/S0036142994277468}

\bibitem{Owen08}
Owen A~B 2008 {\em Ann. Statist.\/} {\bf 36} 2319--2343

\bibitem{MAT98}
Matous\v{e}k J 1998 {\em Journal of Complexity\/} {\bf 14} 527--556

\bibitem{Tezuka03}
Tezuka S and Faure H 2003 {\em Journal of Complexity\/} {\bf 19} 744 -- 757
  ISSN 0885-064X
  \urlprefix\url{http://www.sciencedirect.com/science/article/pii/S0885064X03000359}

\bibitem{SpaMa}
Spanier J and Maize E~H 1994 {\em SIAM Rev.\/} {\bf 36} 18--44 ISSN 0036-1445
  \urlprefix\url{http://dx.doi.org/10.1137/1036002}

\bibitem{Caflisch95}
Caflisch R~E and Moskowitz B 1995 {\em Lecture Notes in Statistics 106\/}
  (Springer-Verlag) pp 1--16

\bibitem{CAF97}
Caflisch R~E, Morokoff W and Owen A 1997 {\em The Journal of Computational
  Finance\/} {\bf 1} 27--46

\bibitem{Blank79}
Blankenbecler R, DeGrand T~A and Sugar R 1980 {\em Phys.Rev.\/} {\bf D21} 1055

\end{thebibliography}

\end{document}